\documentclass[aps, pra, reprint, superscriptaddress]{revtex4-2}

\draft % marks overfull lines with a black rule on the right
\usepackage{graphicx}
\usepackage{dcolumn}
\usepackage{bm}
\usepackage{longtable}
\usepackage[colorlinks=true, allcolors=blue]{hyperref}
\usepackage{amsmath}

\begin{document}

%\preprint{}

\title{Ionization dynamics of submicron-sized clusters in intense ultrafast laser pulses} %Title of paper



\author{Xiaohui Gao}
\email[]{gaoxh@utexas.edu}
%\thanks{}
\affiliation{Department of Physics, Shaoxing University, Shaoxing, Zhejiang 312000, China}

\date{\today}

\begin{abstract}
Submicron-sized targets are found in intense laser-cluster interaction experiments and laser-based material processing.  Here we investigate the internal field localization due to Mie scattering and its effect on ionization dynamics in submicron-sized clusters using Mie calculation and particle-in-cell simulations. As a result of intertwined processes of pulse propagation and ionization, sub-micron nanofocusing dominates at lower intensity and gives rise to an ionization hotspot at the rear of the targets while plasma shielding wins over at a higher intensity, stopping further rear side ionization. As ionization is often the precursor of other processes, understanding the ionization dynamics of ultrafast laser pulses with wavelength-sized nanostructure can be relevant for intense laser-cluster experiments and femtosecond laser micro/nanomachining.
\end{abstract}

\maketitle

\section{Introduction}
Ionization is the first step in the intense laser-matter interaction and is an essential ingredient in many useful phenomena ranging from filamentation~\cite{Couairon2007PR} to high-order harmonic generation~\cite{Brabec2000RoMP}. As a precursor of other important processes such as material modification in ultrashort laser micro/nanomaching~\cite{Gattass2008NP, Mcleod2008NN} and plasma defocusing in intense laser-matter interaction~\cite{Couairon2007PR}, ionization has been under extensive investigation for several decades. Nevertheless, some of the underlying physics are only recently unveiled~\cite{GarciaLechuga2017PRB, Juergens2022AP, Park2022PRL}.

While most study focuses on gases and solids, clusters and other nanostructures, which bridge the gap between gases and solids, possess unique and striking ionization features and attract considerable attention~\cite{Krainov2002PR, Fennel2010RMP, Ostrikov2016RMP,Saalmann2006JPB, Park2022PRL, Gao2019OL}. The ionization threshold of clusters is substantial lower than that of monomers. This feature has been exploited to form plasma waveguide~\cite{Ditmire1998OL} and to characterize the cluster-monomer ratio~\cite{Gao2012APL,Gao2013JAP}. Avalanche ionization can explain the lowered ionization threshold in picosecond pulses~\cite{Ditmire1997PRL}. However, it generally plays little role in sub-100-fs pulses~\cite{Fourment2018PRB}. The increase of the internal field can also be responsible for the lowered ionization threshold. For dielectric spheres with size of tens of microns or more, the refractive focusing increases the internal field, giving rise to reduced ionization threshold~\cite{Efimenko2014JOSAB} and efficient supercontinuum generation~\cite{Favre2002PRL}. In the other limit, as the size decreases to a fraction of the laser wavelength, 
the scattering enters the Rayleigh regime, and the internal field becomes uniform and field enhancement occurs only when there is significant ionization so that the laser frequency approaches the resonant frequency. The calculation includes the avalanche ionization and field enhancement due to Mie resonance shows a ionization threshold similar to that of monomers in femtsecond pulses~\cite{Gao2013JAP}.    

Ionization in sub-micron-sized clusters is relatively unexplored. The refractive focusing breaks down at approximately 10$\lambda$~\cite{Kim2016OE}, where $\lambda$ is the laser wavelength. However, diffractive focusing emerges at a smaller size, leading to the internal or external field enhancement commonly referred as a photonic nanojet~\cite{Chen2004OE}. The sub-wavelength field localization behind spherical particles has found applications in super-resolution imaging~\cite{Darafsheh2012APL} and nanopatterning~\cite{Mcleod2008NN}, and the field localization at the inner rear surface has been exploited for controlled ablation that enables backward jet propulsion of microparticles~\cite{Romodina2022O}. 
A deep understanding of ionization dynamics in this regime is not only of fundamental interest but also of practical importance because it can be useful for the control of ionization in nanostructures and solids for the application of ultrafast laser micro/nanomaching~\cite{Mcleod2008NN}, nano-electroncis~\cite{Suessmann2015NC}, and petahertz electronics~\cite{Garg2016N}.
It is also possible that this effect contributes in part to the lowered ionization threshold observed in Ref.~\onlinecite{Gao2013JAP}. Although the average size is not large there, greater clusters can be present since the size distribution is typically broad~\cite{Mendham2001PRA}. Extremely large clusters may also be found in pulsed gas jets even though the scaling law predicts the size is small~\cite{Li2003CP, Rupp2014JCP}. In addition, our study also has practical significance for laser-cluster applications, as submicron-sized clusters are used for electron acceleration~\cite{Fukuda2007PLA} and ion acceleration~\cite{Fukuda2009PRL, Jinno2018PPCF, Aurand2020NJP}, and the ionization threshold affects the pre-pulse plasma.

In this paper, we numerically investigate the field localization in submicron-sized clusters and the effect of the field localization on the cluster ionization dynamics. The diffractive focusing of submircon-sized clusters initially increases the local field at the rear side of the cluster and creates an ionization hotspot, substantially lowering the ionization threshold. At a higher intensity, the leading part of the pulse ionizes the front surface of the cluster. The resulting overdense plasma layer shields the field, shutting off diffractive focusing effect. 

\section{Mie calculation}
\begin{figure}[htbp]
\centering
\includegraphics[width=0.45\textwidth]{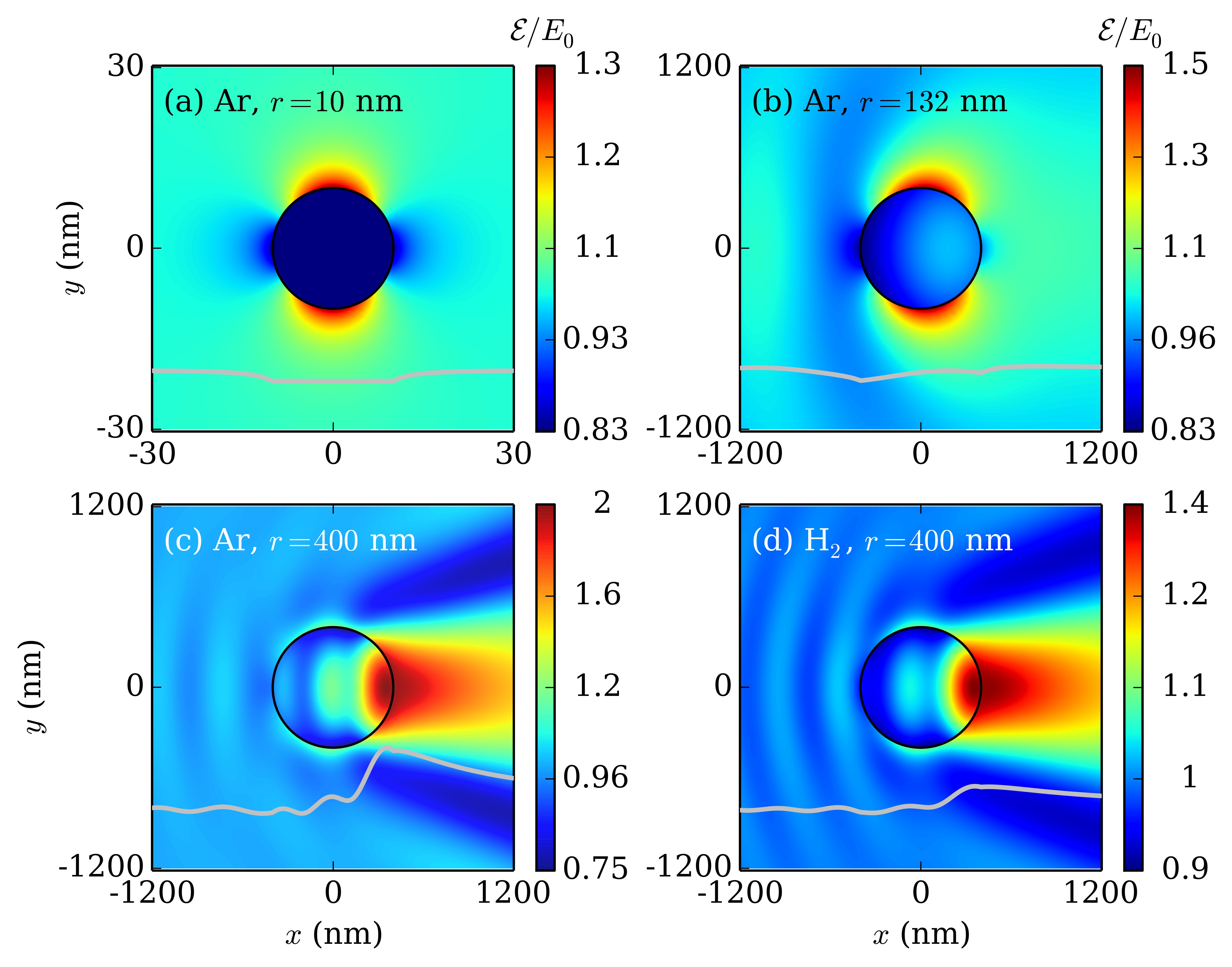}
\caption{Amplitude distribution of the laser field in (a) a 10-nm argon cluster, (b) a 132-nm argon cluster,  (c) a 400-nm argon cluster, (d) a 400-nm hydrogen cluster. The grey lines superimposed in the images are the lineouts at $y=0$.}
\label{f1}
\end{figure}%
Before significant ionization, the fields inside and around a cluster are calculated using a Mie code \textsc{Scattnlay} v2.0~\cite{Ladutenko2017CPC}, which numerically evaluates an analytical solution in terms of infinite series with scattering coefficients. These coefficients are obtained from the Maxwell equations in together with the boundary conditions at the interfaces and are expressed in terms of the size parameter $2\pi r/\lambda$ and the relative refractive index $n_\text{rel}=n_\text{cl}/n_\text{bg}$, where $r$ is the radius of the cluster, $n_\text{cl}$ is the refractive index of the cluster material, and $n_\text{bg}$ is the refractive index of the environment. These calculations are simple and allows rapid parameters scanning. The standard Mie solution of a dielectric sphere is a linear theory which is appropriate when the nonlinear effect such as ionization and Kerr effect can be neglected. In our case, the cluster size is much smaller than the self-focusing distance and Kerr effect is negligible.  

Figures~\ref{f1}(a)-\ref{f1}(c) show the amplitude distribution of the field in argon clusters of three different sizes subject to a 800-nm continuous plane wave linearly polarized in $y$ direction. The amplitude is normalized to $E_0$, which is the peak field in the incident wave. the refractive index of solid argon is taken as $n_\mathrm{Ar}=1.28$. With a 10-nm radius, only the near field outside the cluster is redistributed and the internal field is uniform, as it is in the Rayleigh scattering regime. With a 132-nm radius, the field develops asymmetry along the polarization axis and exhibits a weak local maximum in the cluster. With a 400-nm radius, the light is focused in a hotspot with a size of tens of nanometers at the rear of the cluster. This complicated field distribution can be considered as the superposition of multiple radiation. Similar pattern is found in hydrogen clusters as shown in Fig.~\ref{f1}(d), where $n_{\text{H}_2}=1.11$~\cite{Johns1937CJR} is used.

\begin{figure}[htbp]
\centering
\includegraphics[width=0.45\textwidth]{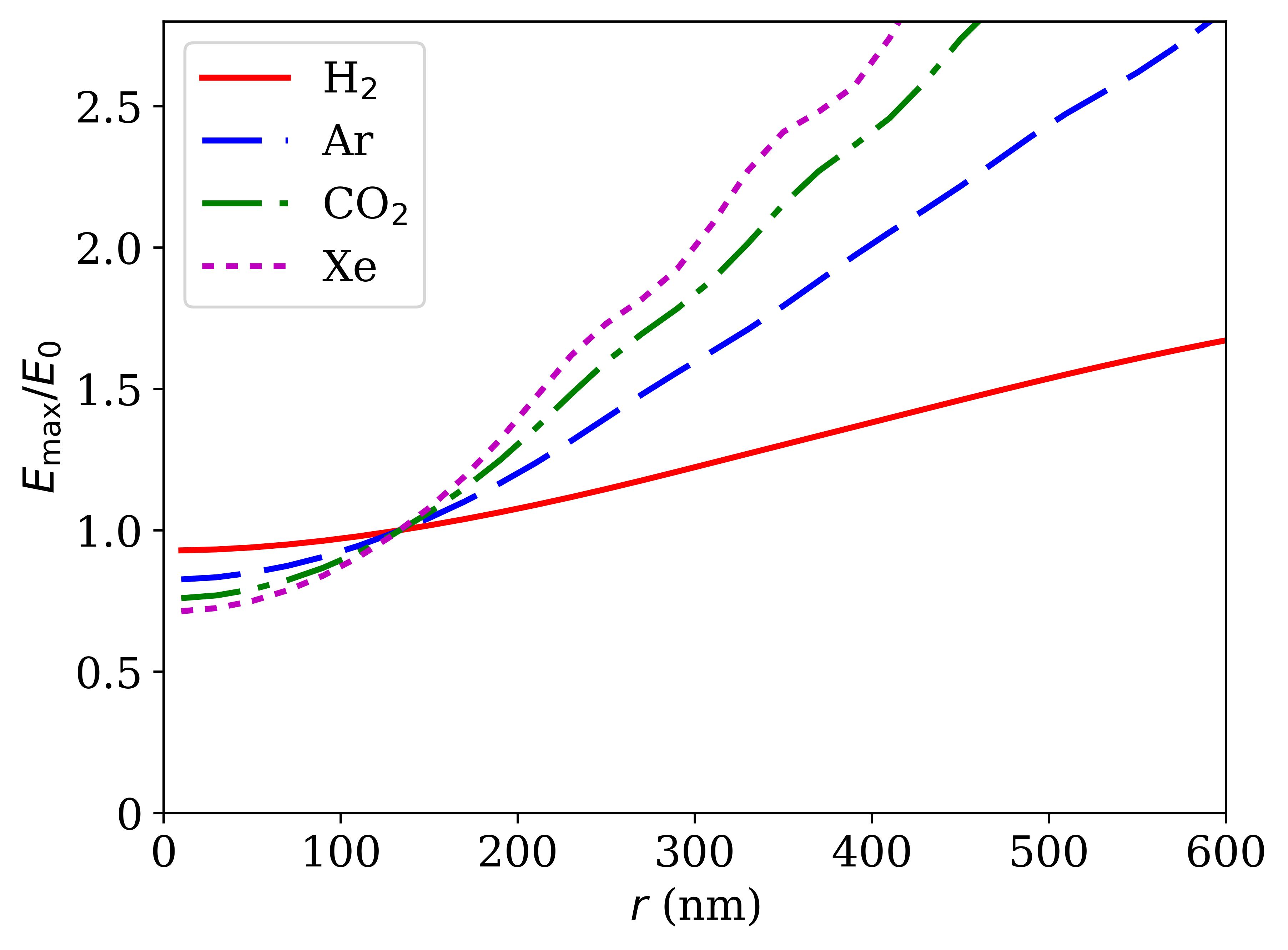}
\caption{Maximum internal field as a function of cluster radius $r$ in a hydrogen cluster (solid red line), an argon cluster (dashed blue line), a CO$_2$ cluster (dash-dotted green line), and a xenon cluster (dotted magenta line).}
\label{f2}
\end{figure}%
The dependence of the internal field enhancement on cluster sizes for commonly used gas species are shown in Fig.~\ref{f2}. The refractive indices of the other cluster material are $n_\mathrm{CO_2}=1.40$~\cite{Loeffler2016AJ} and $n_\text{Xe}=1.49$~\cite{Sinnock1980JPCSSP}. The analytical Mie solution is complicated and the parameter dependence often relies on numerical computation. The maximum internal field reaches the incident value $E_0$ at around $r=130$\,nm. After that, it almost increases linearly with the size. A higher refractive index results in a stronger enhancement. We do not calculate for clusters larger than 600-nm due to limits of numerical precision. As the size increases, the number of digits needs to be increased to calculate of spherical harmonics. 

\section{Particle-in-cell simulations}
When the laser pulse is intense enough to ionize the clusters, the action and reaction of the laser field and plasma give rise to a highly dynamic and nonlinear process, which is not captured by static Mie calculations. To investigate this ionization dynamics, we performed three-dimensional particle-in-cell (3D PIC) simulations using an open-source code \textsc{Smilei}~\cite{Derouillat2018CPC}. This code includes various physics modules such as field ionization, binary collision with collisional ionization (CI).

\begin{figure}[htbp]
\centering
\includegraphics[width=0.45\textwidth]{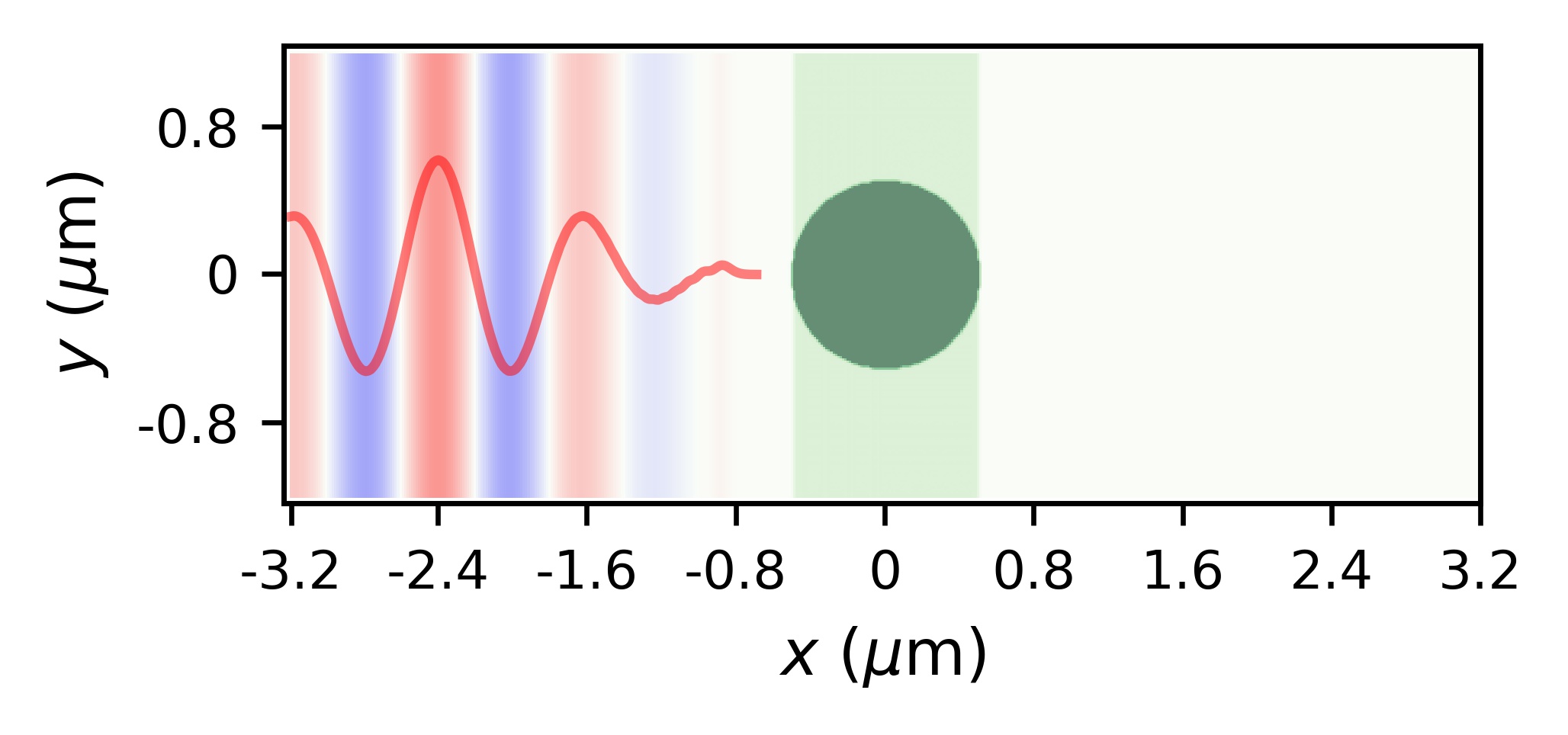}
\caption{Set-up of a typical simulation. The blue and red stripes represent the spatial distribution of the field. The solid red line shows the strength of the field. The dark green sphere denotes the cluster. The light green color denotes an underdense plasma so that there is an index difference between the neutral cluster and the environment to induce an effective diffractive focusing.}
\label{setup}
\end{figure}%
The simulation set-up is presented in Fig.~\ref{setup}. The simulation box was $6.4\times2.4\times2.4\,\mu$m$^3$. The spatial step $\Delta x=\Delta y=\Delta z=\lambda/64$, and the time step $\Delta t$= 0.023 fs, corresponding to a Courant number of 0.956. A laser pulse linearly polarized in the $y$ direction enters the simulation box from the left. The laser wavelength is centered at 800\,nm, and the temporal profile is Gaussian. The full width at half maximum of the field is two cycles. We assume a plane wave with constant profile. 
A submicron-sized hydrogen cluster is located at the center of the simulation box. The time zero $t=0$ marks the moment that the peak of the pulse reaches $x=0$. The number density of hydrogen atoms in the cluster is assumed to be that of the liquid hydrogen, which is $4.56\times10^{22}$\,cm$^{-3}$ or 26.8$n_c$~\cite{Aurand2019PP}, where $n_c$ is the critical density of the plasma at $\lambda=800$\,nm. Each cell in the clusters has 125 macro-particles. Silver-muller boundary conditions are adopted for the electromagnetic fields in the longitudinal direction, and periodic boundary conditions are used for the transverse directions. Quasi-static rates for tunnel ionization are used with Monte-Carlo procedure. Electron-electron collision and electron-ion collision with collisional ionization are included using the binary collision module with electron-ion impact-ionization model. The code does not account for recombination and lowering of ionization potential. As these two effects contribute in the opposite way to the average charge~\cite{Gao2013JAP}, they may get partially canceled. The employed code includes essential physical modules of the cluster interaction with few-cycle pulses, and the convergence of the results is confirmed by reducing the grid size.

Since PIC codes typically consider scenarios where permittivity is dominated by that of free electrons, the index of the neutral atoms is assumed to be unity. 
Thus in normal configuration, the diffractive focusing effect is absent in the code. In the Mie calculation~\cite{Ladutenko2017CPC}, the solution depends on the relative refractive index $n_\mathrm{rel}$. For a hydrogen cluster in vacuum, the relative index is 1.11. In our simulations, we assume a plasma density of 0.2$n_c$ for $-r<x<r$, as indicated by the light green region in Fig.~\ref{setup}. This corresponds to $n_\mathrm{rel}=1.12$. Thus, our simulation gives quantitatively correct result for diffractive focusing before significant ionization. As ionization occurs, the value of $n_\mathrm{rel}$ no longer matches, but its variation would follow the same trend as in the realistic case. The interface of the auxiliary plasmas gives rise to a Fresnel reflection, slightly modifying the field strength. Despite of these simulation flaws, the results remain qualitatively correct. An accurate simulation requires more elaborate PIC capabilities. The simulations are configured in a way that these auxiliary plasmas do not participate in the collisional process.

\begin{figure}[htbp]
\centering
\includegraphics[width=0.45\textwidth]{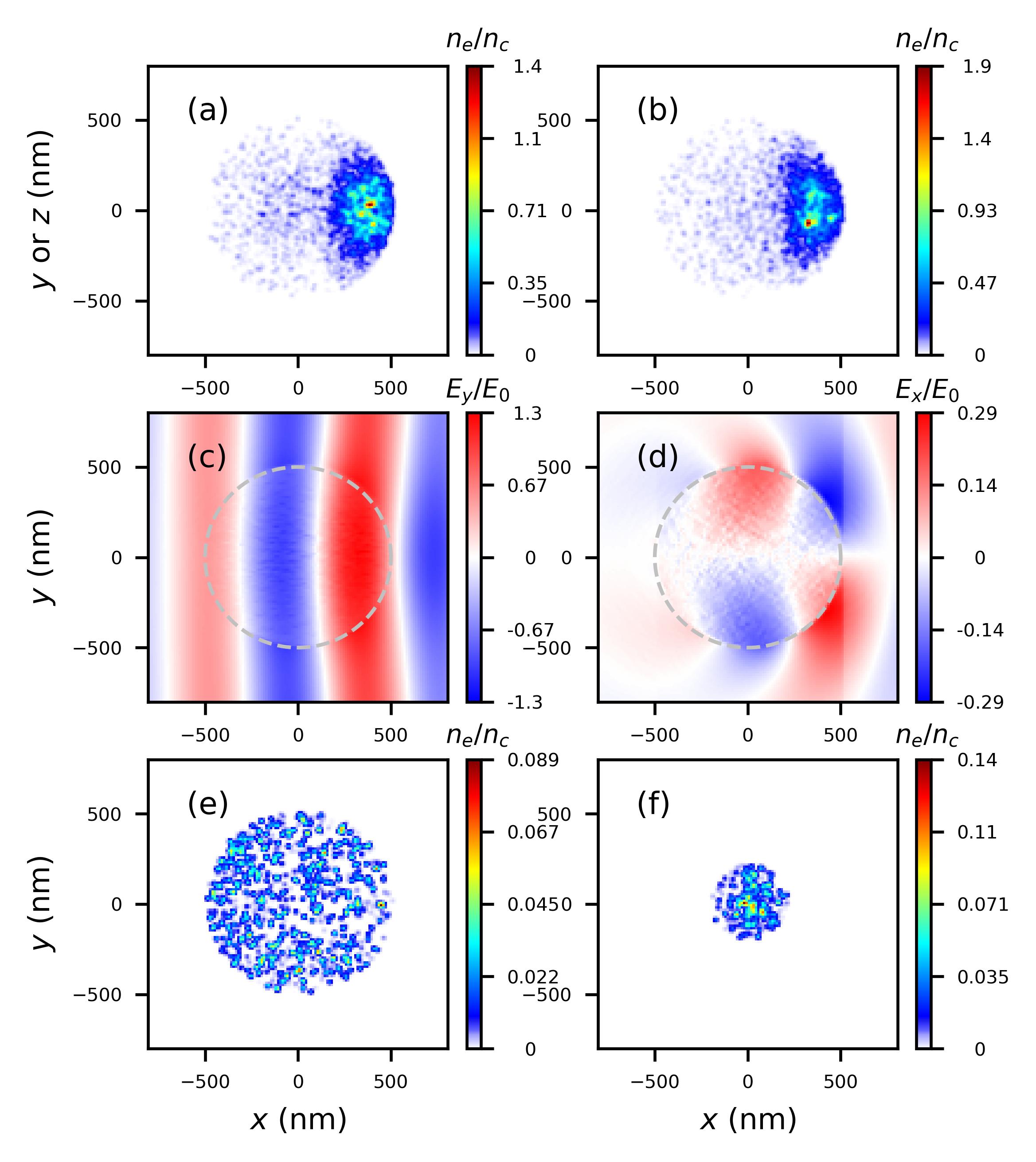}
\caption{Simulations results for $I=7\times10^{13}$ W/cm$^2$. The electron density in the $y$-$x$ plane (a) and $z$-$x$ plane (b) at $t=16$\,fs for a 500-nm cluster. The densities are normalized by the critical density $n_c$. The transverse field $E_y$ map (c) and longitudinal field $E_x$ map (d) in the $y$-$x$ plane at $t=1.0$ fs. The fields are normalized by the incident amplitude $E_0$. (e) The electron density at $t=16$ fs with no index difference between the cluster and the surrounding. (f) The electron density for a 200-nm cluster at $t=16$ fs. }
\label{fig07}
\end{figure}%
Figure~\ref{fig07}(a) and \ref{fig07}(b) show the electron density in two perpendicular planes ($y$,$x$) and ($z$,$x$) at $t=16$\,fs when a 500-nm cluster is irradiated with a pulse of $7\times10^{13}$\,W/cm$^2$ intensity. The colorbar indicates the electron density normalized by the critical density. An ionization hotspot is observed at the rear side of the cluster. The distribution of the transverse field $E_y$ and longitudinal field $E_x$ in the polarization plane at $t=1.0$ fs are given in Fig.~\ref{fig07}(c) and \ref{fig07}(d), respectively. The diffractive focusing of the incident pulse causes the internal maximum of $E_y$ increased by more than 30\% compared with $E_0$. Longitudinal component is also developed due to multiple excitation in Mie scattering, as shown in Fig.~\ref{fig07}(d). Without an index difference between the cluster and the surrounding, uniformly weak ionization is observed as in Fig.~\ref{fig07}(e), and neither field enhancement nor longitudinal field is observed. Figure~\ref{fig07}(f) shows the case with a 200-nm cluster. The diffractive focusing is weaker at a smaller radius. As a result, the ionization is an order of magnitude lower than that in Fig.~\ref{fig07}(a). 

\begin{figure}[htbp]
\centering
\includegraphics[width=0.45\textwidth]{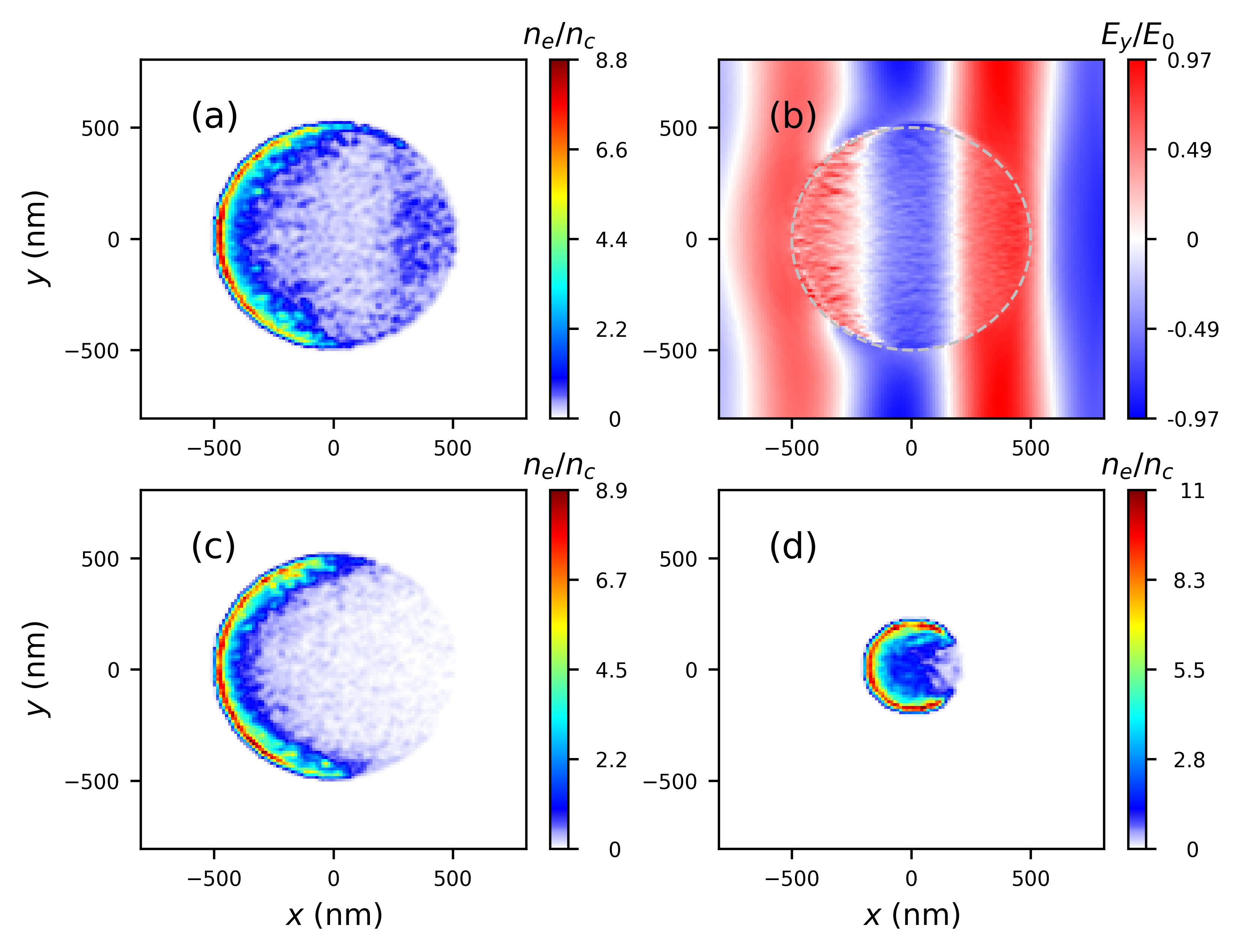}
\caption{Simulation results for $I=2\times10^{14}$ W/cm$^2$. (a) The electron density in the polarization plane for a 500-nm cluster at $t=16$ fs. (b) $E_y$ in the polarization plane at $t=1.0$ fs. (c) The electron density at $t=16$ fs with no index difference between the cluster and the surrounding. (d) The electron density for 200-nm cluster at $t=16$ fs.}
\label{figI2}
\end{figure}%
The ionization maximum occurs at the rear of the cluster only for a limited range of intensity near the threshold. At a higher intensity, the front surface becomes heavily ionized. The electron density and field distribution for a 500-nm cluster subject to $I=2\times10^{14}$ W/cm$^2$ are presented in Fig.~\ref{figI2} (a) and \ref{figI2}(b), respectively. At this intensity, the leading part of the pulse is intense enough to create sufficient ionization at the front surface. Due to the skin effect, this ionization is limited to a thin layer, and the internal field is shielded as demonstrated in Fig.~\ref{figI2}(b). Figure~\ref{figI2}(c) shows the electron density without the diffractve focusing. The overall ionization pattern is similar, except for the minor feature at the rear of the cluster. Figure~\ref{figI2}(d) shows the case of a 200-nm cluster, where a higher plasma density is observed. 
Other particle-in-cell simulations of the femtosecond laser interaction with micron-sized target shows that the ionization and heating is limited to the skin depth in helium droplets~\cite{Liseykina2013PRL} and hydrogen droplets~\cite{Zastrau2015JPBAMOP}, which are consistent with our results. Similar scenario is also observed in experiments, where the average radius of CO$_2$ clusters in the wake of the prepulse is reduced from 0.18 $\mu$m to 0.15 $\mu$m~\cite{Faenov2012LPB}.

\begin{figure}[htbp]
\centering
\includegraphics[width=0.45\textwidth]{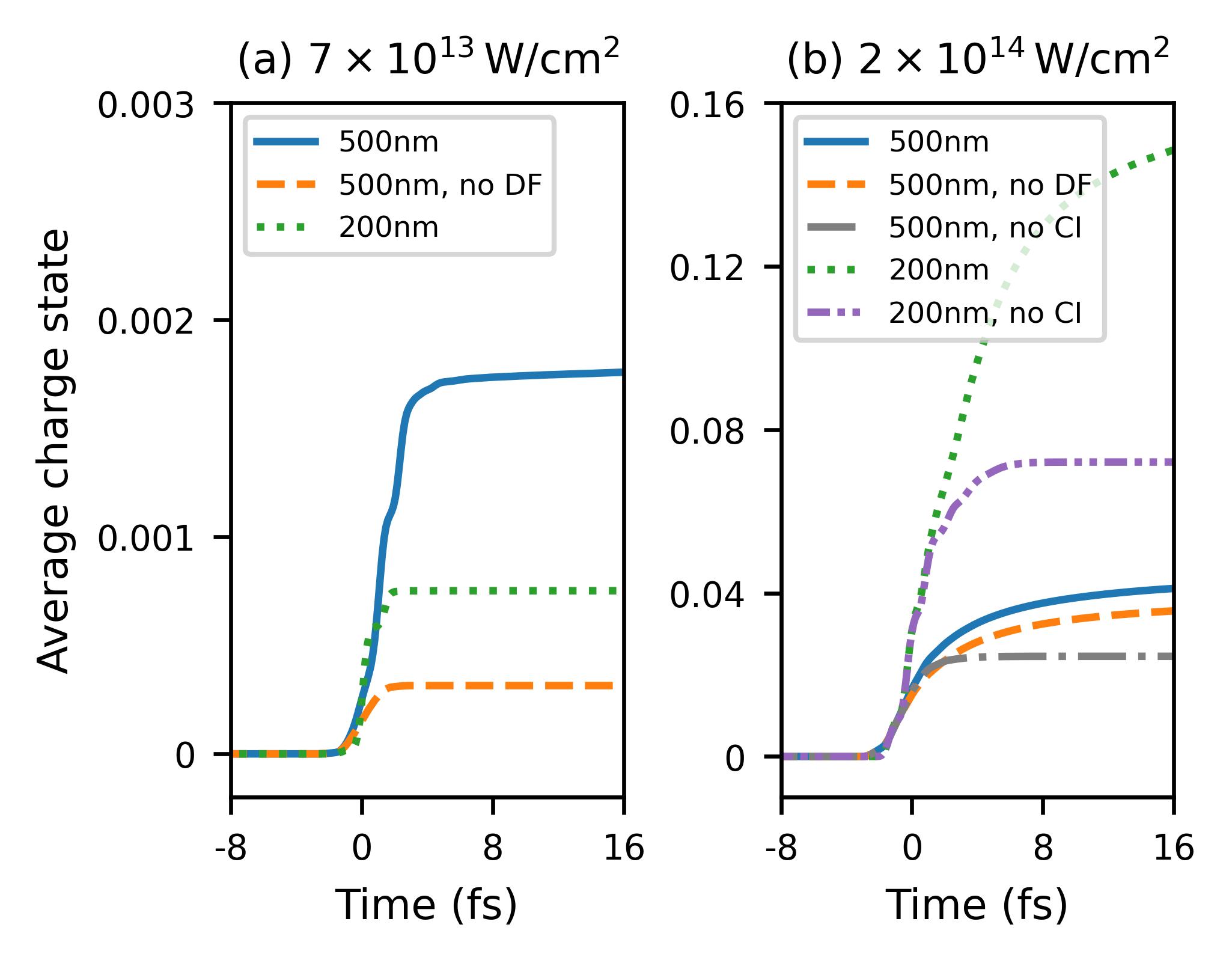}
\caption{(a) Time evolution of the average charge in the cluster for $I=7\times10^{13}\,$W/cm$^2$ with a 500-nm cluster (solid blue line), a 500-nm cluster with no index contrast (dashed orange line), and a 200-nm cluster (dotted green line) (b) The cases for $I=2\times10^{14}\,$W/cm$^2$. The dashdotted gray line represents the case for a 500-nm cluster without collisional ionization, and the dashdotdotted purple line represents the case for a 200-nm cluster without collisional ionization.}
\label{fig1d}
\end{figure}%
Figures~\ref{fig1d}(a) and \ref{fig1d}(b) present the time evolution of the volume-averaged charge state at $7\times10^{13}\,$W/cm$^2$ and $2\times10^{14}\,$W/cm$^2$, respectively. Near the ionization threshold, a larger size causes stronger diffractive focusing and is beneficial for ionization. At a higher intensity, however, a smaller size leads to a higher ionization. This is because the ratio of atoms in the skin layer increases with a decreasing size. In addition, the charge density at the front surface is higher as shown in Fig.~\ref{figI2}. Note that the ionization continues after the pulse is gone. This rise is mainly due to collisional ionization. The dashdotted gray line and the dashdotdotted purple line shows the cases for a 500-nm cluster and a 200-nm cluster, respectively, when the collisional ionization is switched off. Here the elastic collision remains. Thee charge state levels off quickly.  

\section{Summary}
We have shown that sub-micron nanofocusing of large clusters reduces the ionization threshold. This diffractive focusing effect shows a strong dependence on the size and intensity, as confirmed by our PIC simulations. At a higher intensity, however, the diffractive focusing only causes a minor difference in the ionization due to the shielding of the plasma layer at the front surface. The intertwined processes of pulse propagation and ionization manifest themselves as diffractive focusing at low intensity and plasma shielding at a higher intensity. Control and optimization of the ionization dynamics may find applications in intense laser-cluster experiments such as preplasma control in particle acceleration and ionization control of dust in mid-infrared filamentation~\cite{Woodbury2020PRL,Zhang2019OE,Rudenko2020O}. It also facilitates our understanding and control of ionization and heating phenomena in nanostructures and solids for the application of laser-based material processing and petahertz electronics.

\begin{acknowledgments}
This work was supported by Natural Science Foundation of Zhejiang Province (grant LY19A040005).
\end{acknowledgments}

%\bibliographystyle{aipnum4-2}

%aipnum4-2.bst 2019-01-14 (MD) hand-edited version of apsrev4-1.bst

\end{document}